# Extensive air showers' arrival direction distribution by the TSU array under GELATICA experiment

## Yu. G. Verbetsky[1], M. S. Svanidze[1], A. Iashvili[1], I. Iashvili[2], L. Kakabadze[1]

1. E Andronikashvili Institute of Physics under Tbilisi State University, Tbilisi, Georgia
2. The State University of New York at Buffalo, USA

**Email address**: yuverbetsky@mail.ru (Yu. G. Verbetsky), mananasvanidze@yahoo.com (M. S. Svanidze)

**Abstract**: The distribution of the arrival zenith angle of the Extensive Air Showers (EAS) with a wide range of a total number of charged particles is studied using the experimental data obtained using the EAS 4-detector array "TSU" in Tbilisi. The station is a part of the GELATICA net in Georgia (GEorgian Large-area Angle and TIme Coincidence Array), which is devoted to the study of possible correlations in the arrival times and directions of separate EAS events over large distances. It is shown that the distribution function with the conventional exponential dependence of showers' flux on absorbing air thickness provides a good approximation for the arrival direction distribution. The dependence of the EAS absorption path estimation on the angular trimming boundary of data set is studied; the necessity of strict verification of the used value of data trimming boundary is stated.

## *Introduction*

Extensive Air Showers (EAS) development in the atmosphere with accompanying absorption manifests itself through the arrival direction distribution. That is why an interest to such investigations is long-standing [1-8]. The distribution of the zenith angle $\theta$ of the shower arrival direction is usually studied under the assumption of azimuth isotropy for both the Cosmic Ray phenomenon and the measuring equipment.

It has been shown previously [6] that the distribution of the zenith angle weakly depends on the energy of the Primary Cosmic Ray particles. This feature makes it possible to investigate the subject using small installations incapable of directly measuring of EAS energy. The data discussed hereafter is obtained by a small installation (EAS goniometer "TSU") arranged under the iron roof in the second building of Javakhishvili Tbilisi State University. The station is a part of the GELATICA net in Georgia (GEorgian Large-area Angle and TIme Coincidence Array) [9-11] and this long-term experiment is devoted to the study of possible correlations between the separate EAS events over large distances [12] by their arrival times and directions – the so-cold "super-preshowers" [13].

## *1. Description of the installation*

The TSU installation is situated at the geographical location (41.710439° N, 44.776981° E) and sits at an altitude of $h_{TSU} = (474.5 \pm 2.5)m$ by GPS estimation. The installation includes four scintillator detectors located under the standard (0.5 mm) iron roof, which are controlled by the data acquisition (DAQ) card [14] operating under PC control with a LabView interface for Windows. Detectors are arranged (Figure 1) approximately in the corners of a square with side $a \approx 10 m$. Each detector of the installation consists of a 5 cm thick scintillator slab with an area of $(50 \times 50) cm^2$ supplied with a photo-multiplier tube (PMT). The PMT pulses, initiated by the passage of EAS charged particles through the scintillator material, are read by the DAQ card.

The equipment measures the pulse delay relative to the 4-fold pulse coincidence with a $\tau = 1.25 ns$ time slicing step. The data are stored on the PC as integer values $k_0, k_1, k_2, k_3$, equal to the numbers of delay slices for the respective detectors. This information allows a posterior estimation [15] of the direction of the local tangent plane of the arriving EAS front imagined surface.

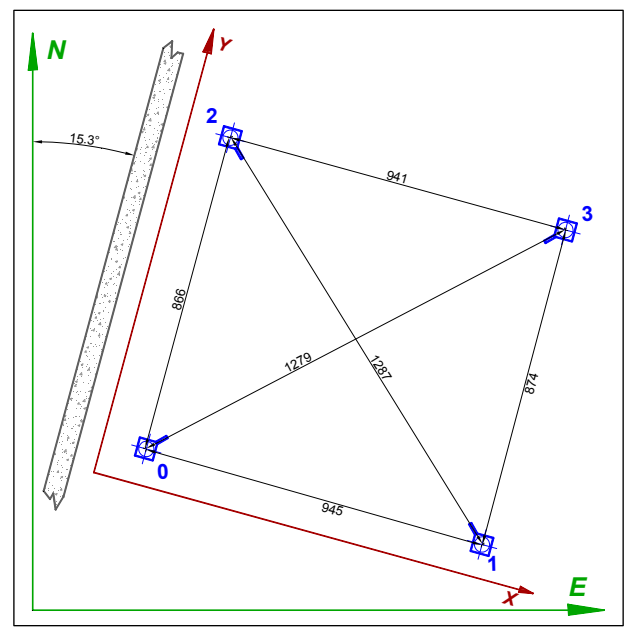

**Figure 1** The TSU array layout. The gray strip displays the horizontal profile of the wall in the roof space used; the numerals **0, 1, 2, 3** are the labels of the detectors.
Dimensions are measured in centimeters.
The XY and East-North reference frames are shown.

## 2. Direction estimation by square EAS goniometer

It is practical to estimate the local direction of the EAS front arrival by the unit directing 3-vector (ort) **N** of the front's local tangent plane. This assumption is approximately correct on average. The components of **N** are the direction cosines with respect to the employed rectangular coordinate system. It is assumed that the front of the shower is moving with light velocity $c$.

The TSU installation is a planar EAS goniometer, permitting the linear estimation of only the planar (horizontal) components of the directing ort **N** of the EAS front's local tangent plane [15].

For the very special case of the detectors' disposition in the corners of a square, the estimation of the horizontal 2D projection of this ort is

$$\mathbf{n} = \begin{pmatrix} n_x \\ n_y \end{pmatrix} = \frac{c\tau}{2a} \begin{pmatrix} (k_1 + k_3) - (k_0 + k_2) \\ (k_2 + k_3) - (k_0 + k_1) \end{pmatrix} \quad (2.1)$$

with respect to the XY reference frame (Figure 1). The estimation of the main dispersion of this 2D-vector components', due to the time fluctuation of the particles' passage through the detectors, is

$$\sigma_x^2 = \sigma_y^2 = \left(\frac{c\tau}{2a}\right)^2 \left((k_0 + k_3) - (k_1 + k_2)\right)^2,$$

while the correlation vanishes. Only statistical uncertainty is taken into account. Naturally, the real coordinates of the detectors, not the square design approximation, are used in the practical calculations. (The exact expressions [15] are simple but cumbersome.)

Certainly, there exist some else sources of fluctuation of ort components' estimation, i.e. variation of the passage position of the triggering particle in every detector slab, uncertainty of the detectors' locations measurements, etc. The corresponding dispersions prove to be considerably less important then the received main one. Yet, these additional dispersion matrixes are still applied to the processing of the TSU installation data.

It is obvious that the measured values of the direction ort components (2.1) (for a square-plan goniometer) possess the magnitudes on the square lattice with a step $c\tau/2a$ due to the integer values $k$ of the delay slice numbers. That is why any EAS event corresponds to only one of the lattice sites on the $(n_x, n_y)$ plane, representing some separate area of possible directions on the celestial hemisphere, and the set of these site neighborhoods become the natural bins of the 2D histogram. The idea is approximately usable for real TSU goniometer, as the squareness of the detectors' positions in Figure 1 is violated weakly.

The measured 2D-distribution of the arrival directions of 21648 EAS events recorded by the TSU goniometer (with an average rate near 20 events / hour) is shown in Figure 2. This event number histogram visually represents the data analyzed hereafter. The rough axial symmetry is evident.

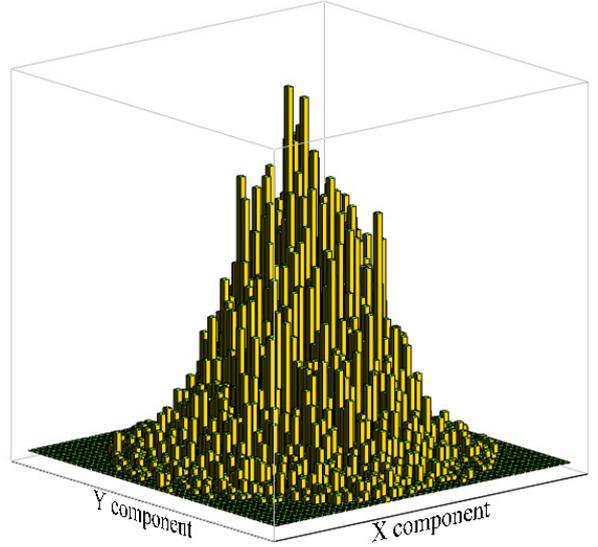

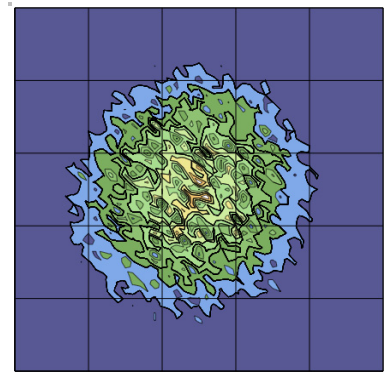

**Figure 2** The histogram of $(n_x, n_y)$ components of EAS arrival direction orts by the TSU goniometer data in the XY reference frame.
The "topographic map" of the histogram visually demonstrates a rough axial symmetry of measured arrival directions.

## 3. Description of the atmosphere mass thickness

The consistent investigation of the EAS arrival directions' distribution as a final goal of this study needs a reliable description of the EAS absorption by the air surrounding the installation, i.e. some reasonable model of the atmosphere. The most reliable model is the one used by the International Civil Aviation Organization [16] (ICAO). This standard model allows the calculation of the air mass thickness $X^{(ICAO)}(h, \theta)$ above a given altitude $h$ in the direction along the zenith angle $\theta$ using the integration of a piecewise smooth function. This model is tenable but awkward for further calculations with subsequent numerical integration. Therefore the analytical approximation is required.

At any altitude the air mass thickness grows as the direction ort approaches the horizon, remaining restricted. It is convenient to represent this air mass thickness by the common expression:

$$X(h, \theta) = X^{\uparrow}(h) \cdot U(\theta, h), \quad (3.1)$$
$$1 \leq U(\theta, h) < \infty; \quad U(0, h) = 1.$$

Here $X^{\uparrow}(h) = X(h,0)$ is the vertical mass thickness of the air at the altitude $h$, while the normalized air thickness function $U(\theta,h)$ describes the angular dependence of the thickness at this altitude. Apparently, the last nondimensional function monotonically increases with the zenith angle value, is bounded in the horizon limit, and it turns into unit in the zenith direction.

The most useful is the conventional model of the flat atmosphere (FAM), where
$$U^{(flat)}(\theta,h) = \sec(\theta). \quad (3.2)$$

Here, the air mass thickness calculated in accordance with the last expression grows unrestrictedly in the horizon vicinity. Only directions within the 60° limit of zenith angle are usually allowed in this model for cosmic radiation absorption studies.

That is why somewhat more sophisticated model of spheric atmosphere (SAM) [8] is more suitable. It calls for ordinary geometric calculation to get the dependence
$$U^{(spher)}(\theta,h) = \frac{1+(1+C(h))}{\cos(\theta)+\sqrt{(1+C(h))^2-\sin^2(\theta)}} \quad (3.3)$$

for the atmosphere imagined as a spheric layer of the air with a limited vertical depth. Here the specific parameter function $C(h)$ describes the influence of the fictitious atmosphere height above the concerned point altitude in relation with the terrestrial globe dimensions. That is why the value of this parameter at any altitude must be estimated by the best possible matching of (3.3) with the standard ICAO atmosphere model. The SAM air mass thickness in the limit $C(h) \to 0$ (i.e. as if the radius of the globe tends conditionally to infinity) tends to the form (3.2) of the FAM.

For the TSU goniometer the known altitude $h_{TSU} = (474.5 \pm 2.5)m$ allows the calculation of both SAM parameters needed:
$$X^{\uparrow}_{TSU} = X^{(ICAO)}(h_{TSU},0°) = (978.8 \pm 0.3) \ g/cm^2 \\ C_{TSU} = C(h_{TSU}) = (2054.3 \pm 0.1) \cdot 10^{-6}. \quad (3.4)$$

The comparison of the three models in the horizon vicinity is shown in Figure 3. The SAM is satisfactory with a maximal deviation of 0.5% from the ICAO model for zenith angles $0 < \theta < 87°$ in the case of the TSU installation. The FAM becomes unacceptable for much lower zenith angles. In contrast to the FAM, the SAM underestimates the total air mass thickness in the horizontal direction.

We note that any regular model of air mass thickness becomes unsuitable in the horizon vicinity both due to the matter surrounding the installation and on account of the relief of the land, so the applicability limit of 87° for the SAM overlaps any need.

The spheric atmosphere model ((3.1),(3.3)) with parameters values (3.4) will be used for numerical calculations only. The common form (3.1) is sufficient for the following definitions.

## 4. Fundamental distribution of the EAS arrival directions

We shall assume that all EAS developed in the atmosphere are absorbed at low altitudes in compliance with the usual exponential rule [1-8]. Thus the flux density of the EAS observed in the solid angle differential $\sin(\theta)\,d\theta d\varphi$ after propagation through the air depth $X(h,\theta)$ is proportional to
$$\exp\left(-\frac{X(h,\theta)}{\Lambda}\right) \cdot \sin(\theta)\,d\theta d\varphi.$$

Here $\Lambda$ is the EAS absorption path required. Taking into consideration that the TSU goniometer employs the flat detectors located in the horizontal plane (i.e. adding a $\cos(\theta)$ factor to the expression above), let us integrate the obtained flux expression by the azimuth to get a zenith angle distribution in the form of
$$f_\theta(\theta|\Lambda,h) \propto \sin(\theta)\cos(\theta) \cdot e^{-\frac{X^{\uparrow}(h)}{\Lambda} \cdot U(\theta,h)}. \quad (4.1)$$

Hereinafter the symbol "$\propto$" stands for "*equal accurate to a normalization factor*", used for definitions of the functional forms of distributions. The air mass normalized thickness function $U(\theta,h)$ (3.1) is defined previously.

As mentioned above, the planar goniometers are capable of a straight estimation of two components of the EAS arrival direction unit vector only, i.e. $(n_x, n_y)$, being parallel to the detectors' location plane [15]. That is why the immediate variable, independent of any additional assumption and measuring the event direction separation from the zenith direction, is the estimated

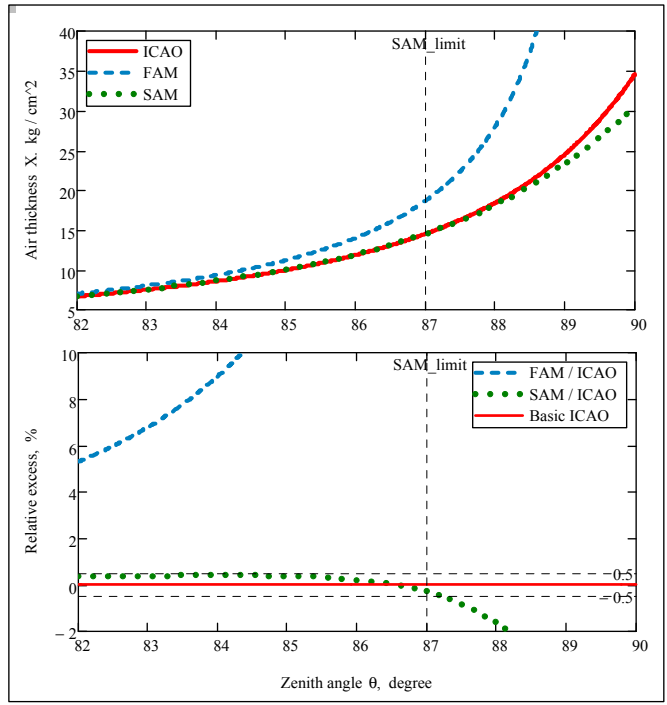

**Figure 3** Comparison of three models of air mass thickness for TSU location.

length of the direction vector projection onto the detectors' plane

$$\beta = \sqrt{n_x^2 + n_y^2}. \quad (4.2)$$

This variable is an indirect estimate of the usual zenith angle. The corresponding geometric zenith separation variable

$$\alpha = \sin(\theta)$$

is restricted to the finite interval $0 \leq \alpha \leq 1$, while the directly estimated value $\beta$ of the event's zenith separation may exceed the geometric limit of unity due to estimation errors.

Let us express the zenith angle distribution (4.1) via the true zenith separation variable $\alpha$; as $d\alpha = \cos(\theta)\,d\theta$, the fundamental distribution of true zenith separations gets the form:

$$f_\alpha(\alpha|q,h) \propto \alpha \cdot \exp\{-q(h,\Lambda)\cdot V(h,\alpha)\};$$
$$q = X^\uparrow(h)/\Lambda; \quad U(h,\theta) \equiv V(h,\sin(\theta)). \quad (4.3)$$

Here the value $q$ is the EAS absorption range number in the vertical direction; the function $V(h,\alpha)$ is the same normalized thickness function (3.1) expressed via the true zenith separation $\alpha$. The complete normalized fundamental distribution of the EAS arrival true zenith separations has the form:

$$f_\alpha(\alpha|q,h) = \frac{\alpha \cdot e^{-q(h,\Lambda)\cdot V(h,\alpha)}}{\int_0^1 e^{-q(h,\Lambda)\cdot V(h,\alpha)}\,\alpha\,d\alpha} \Theta(\alpha)\Theta(1-\alpha) \quad (4.4)$$

Here $\Theta(\alpha)$ is the ordinary Heaviside unit step function.

The EAS absorption range number parameter $q(h,\Lambda)$ is a combination of both the position of the goniometer (through the vertical air mass thickness $X^\uparrow(h)$) and the EAS propagation in the atmosphere – through the EAS absorption path $\Lambda$. Our immediate task is to estimate the last absorption path on the grounds of the obtained TSU data of the measured zenith separations $\beta$ of the EAS events' set.

## 5. Resolution function

For the estimation of the parameters of a fundamental distribution on basis of some experimental data it is necessary to take into account the existing distortion of the fundamental distribution by the measurement errors. That is why it becomes necessary to compare the existing data with the distribution distorted by the resolution function dependent on the errors' distribution.

The detectors of the TSU installation are located almost symmetrically in the vertices of a square (Figure 1). The estimations of the components of the EAS arrival direction vector are almost uncorrelated and equal-dispersion in this case. The components' estimations are obtained by means of a linear transformation (like the (2.1) expression) of directly measured random timing $k$ numbers of signals' from the detectors. Therefore it is possible to use the assumption that the joint distribution of the estimates of $(n_x, n_y)$ components can be approximated by the general Normal 2D-distribution

$$G(\mathbf{n}|\mathbf{n}_0, \mathfrak{D}) \propto \exp\left\{-\frac{1}{2}(\mathbf{n}-\mathbf{n}_0)^T \cdot \mathfrak{D}^{-1} \cdot (\mathbf{n}-\mathbf{n}_0)\right\};$$
$$\mathbf{n}_0 = \begin{pmatrix} n_{x,0} \\ n_{y,0} \end{pmatrix}; \quad \mathbf{n} = \begin{pmatrix} n_x \\ n_y \end{pmatrix}; \quad \mathfrak{D} = \begin{pmatrix} \sigma_x^2 & \rho\sigma_x\sigma_y \\ \rho\sigma_x\sigma_y & \sigma_y^2 \end{pmatrix}. \quad (5.1)$$

Here the vector $\mathbf{n}_0$ represents the unknown true 2D-direction. Only the vector $\mathbf{n}$ may be measured, with the dispersion matrix $\mathfrak{D}$ of vector components, of course.

Since the positions of the detectors in Figure 1 are approximately axially symmetric, most of measured correlation coefficients $\rho$ of the $(n_x, n_y)$-components' estimations are negligibly small. (Correlation coefficient value for the TSU goniometer data varies near the $\bar{\rho} = 0.048$ and almost does not depend on the $\beta$ value). That is why we dare to replace the exact dispersion matrix $\mathfrak{D}$ with the identity-proportional one:

$$\mathfrak{D} \Rightarrow \sigma^2 \mathfrak{I}$$

with an equivalent dispersion $\sigma^2$ defined by

$$\sigma^2 = \sqrt{|\mathfrak{D}|} = \sqrt{\sigma_x^2 \sigma_y^2 (1-\rho^2)}. \quad (5.2)$$

The determinants of both matrixes are equal as a consequence of this definition. The averaged value of the equivalent dispersion $\sigma^2$ for the TSU goniometer is defined further.

The possibility of this replacement is another advantage to using azimuthally symmetric goniometers.

Let us express the symmetrized distribution (5.1) in a polar coordinate system with radius $\alpha$ for a true vector $\mathbf{n}_0$ and radius $\beta$ for a measured vector $\mathbf{n}$. This notation allows us to integrate the simplified Normal 2D-distribution by the azimuth – to obtain [8] the needed radial distribution of the measured zenith separation $\beta$. Consequently the resolution function can be defined as the conditional distribution of the unbounded measured separation $\beta$ estimation under the assumption that $\alpha$ is the known true value of this separation:

$$\mathrm{Rr}(\beta|\alpha) \propto \beta \cdot \exp\left\{-\frac{(\alpha-\beta)^2}{2\sigma(\beta)^2}\right\} \cdot I_0^*\left(\frac{\alpha\beta}{\sigma(\beta)^2}\right); \quad (5.3)$$
$$\beta \in [0,\infty); \quad \alpha \in [0,1].$$

Here the scaled modified Bessel function $I_0^*(x) = e^{-x} I_0(x)$ is defined to get a suitable form of the expression.

The set of the processed EAS observation data obtained by the TSU goniometer contains both the estimations of $(n_x, n_y)$-components of horizontal projections of the arrival direction vectors and the complete dispersion matrixes $\mathfrak{D}$ for every observed event. So the measured separation $\beta$ (4.2) and the equivalent dispersion $\sigma^2(\beta)$ (5.2) are ascertained for all events. The dependence of the last dispersion on the $\beta$ separation is shown in Figure 4 together with the respective regression polynomial. There is no need to reveal this function explicitly.

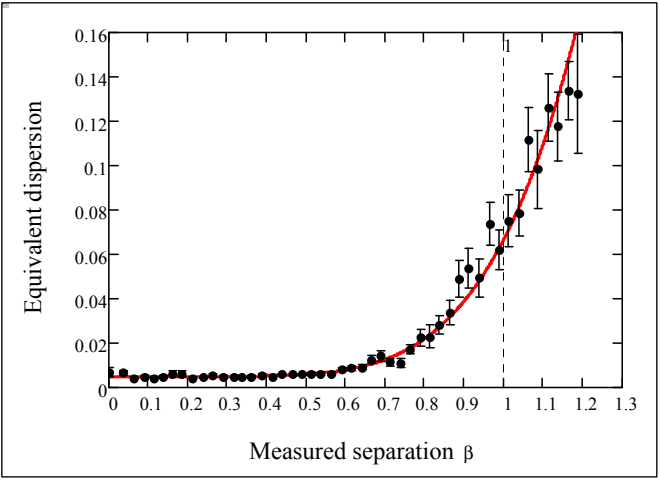

**Figure 4** Equivalent dispersions' dependence on the measured zenith separation β is shown as provided by the TSU data. The regression polynomial is shown in red.

Now the resolution function $\mathrm{Rr}(\beta|\alpha)$ (5.3) is completely defined for the TSU goniometer circumstances under the usual normalizing requirement

$$\int_0^\infty \mathrm{Rr}(\beta|\alpha)\,d\beta = 1$$

for the conditional distributions.

The characteristic slices of this (normalized) resolution function are shown in Figure 5 for several values of the true zenith separation $\alpha$ listed in the picture. Note that the resolution function creates (due to the measurement errors) some number of events with measured separations $\beta$ beyond the horizon limit.

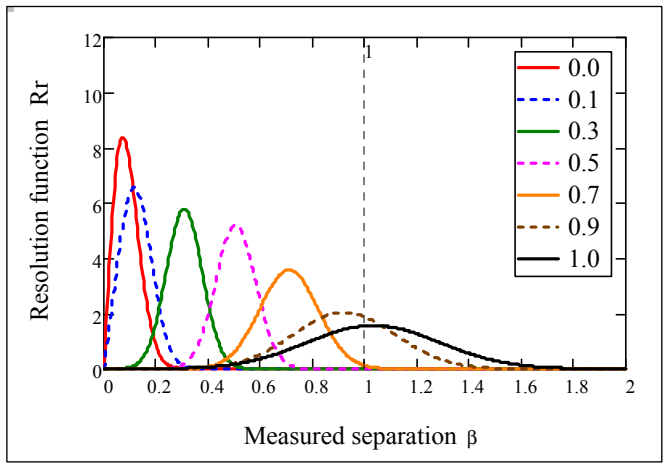

**Figure 5** The resolution functions $\mathrm{Rr}(\beta|\alpha)$ for listed values of the true zenith separations $\alpha$.

## 6. Distribution of the measured zenith separations

It is conventional to design a probability density function of a possible measurement of some zenith separation $\beta$ value by means of averaging of the resolution function $\mathrm{Rr}(\beta|\alpha)$ (5.3) over the fundamental distribution of the nonmeasurable values of true zenith separations (4.3):

$$f_\beta(\beta|q) \propto \int_0^1 \mathrm{Rr}(\beta|\alpha) \cdot f_\alpha(\alpha|q)\,d\alpha. \quad (6.1)$$

Hereinafter the altitude argument $h$ is omitted as all calculations are devoted to the specified installation at fixed altitude.

It is timely to put into operation the normalization factor for the distribution under consideration. Let us take into account that we need the distributions defined on the bounded segment $\beta \in [0, B] \subset [0, \infty)$ of the $\beta$ variable and define the normalizing function as the interval integral of the distribution (6.1) over this segment:

$$\int_0^B f_\beta(\beta|q)\,d\beta = \int_0^1 \mathrm{Ri}(\alpha|B) \cdot f_\alpha(\alpha|q)\,d\alpha.$$

Here the integral resolution function is defined as:

$$\mathrm{Ri}(B|\alpha) \propto \int_0^B \mathrm{Rr}(\beta|\alpha)\,d\beta.$$

So the normalized distribution of the measured zenith separation over the bounded $\beta$-segment is defined as:

$$f_\beta(\beta|q,B) = \frac{\int_0^1 \mathrm{Rr}(\beta|\alpha) \cdot f_\alpha(\alpha|q)\,d\alpha}{\int_0^1 \mathrm{Ri}(B|\alpha) \cdot f_\alpha(\alpha|q)\,d\alpha} \cdot \Theta(\beta)\Theta(B-\beta). \quad (6.2)$$

This distorted distribution is used for the comparison with data and the definition of the likelihood function for the step-by-step estimations of the parameter $q$ for several values of trimming boundaries $B$.

## 7. The maximal likelihood equation for the parameter $q$ estimation

Let us consider the measured separations data sample as an order statistics of the total sample size $N_\mathrm{T}$:

$$\{\beta\} = \left\{ 0 \leq \beta_1 \ldots \leq \beta_j \ldots \leq \beta_{N_\mathrm{T}},\ j = 1 \ldots N_\mathrm{T} - 1 \right\}.$$

The size of trimmed subsample $0 \leq \beta_j \leq B$ is defined as $N_B = \sum_{j=1}^{N_\mathrm{T}} \Theta(B - \beta_j)$; here the trimming boundary $B$ is a free parameter.

The logarithm of the likelihood function for the distribution of measured zenith separations (6.2) with respect to the last trimmed subsample is:

$$\mathfrak{L}(q|\{\beta\},B) = \sum_{j=1}^{N_B} \ln\left(f_\beta(\beta_j|q,B)\right).$$

The common maximal likelihood equation $d\mathfrak{L}(q|\{\beta\},B)/dq = 0$ for the known form (6.2) of the bounded distorted distribution allows the explicit form of the equation for the parameter $q$ estimation (under $\beta \leq B$ restriction):

$$\mathrm{Vi}_1(q|B) = \frac{1}{N_B}\sum_{j=1}^{N_B} \mathrm{Vr}_1(q|\beta_j). \quad (7.1)$$

Here two average functions of the normalized air thickness function $V(\alpha)$ (4.3) are used:

$$\mathrm{Vr}_1(q|\beta) = \frac{\int_0^1 \mathrm{Rr}(\beta|\xi)\cdot f_\alpha(\xi|q)\cdot V(\xi)\,d\xi}{\int_0^1 \mathrm{Rr}(\beta|\xi)\cdot f_\alpha(\xi|q)\,d\xi}, \quad (7.2)$$

(it depends on the measured separation $\beta$ value);

$$\mathrm{Vi}_1(q|B) = \frac{\int_0^1 \mathrm{Ri}(B|\xi)\cdot f_\alpha(\xi|q)\cdot V(\xi)\,d\xi}{\int_0^1 \mathrm{Ri}(B|\xi)\cdot f_\alpha(\xi|q)\,d\xi}, \quad (7.3)$$

(it depends on the sample trimming boundary $B$).

The maximal likelihood equation (7.1) has the form of equalization for two types of averages of the normalized air thickness.

On the left side of the equation the $\mathrm{Vi}_1(q|B)$ function (7.3) does not depend on the data sample $\{\beta\}$, – it is simply a function of the parameter $q$ and boundary value $B$. The function on the right-hand side of the equation is a sample-mean of averages (7.2) depending on the parameter $q$ too. Both expressions depend implicitly on the accepted form of the fundamental distribution (4.3) and on the atmosphere model used.

The equation (7.1) has to be solved numerically. The solution definitely depends on the data sample and trimming boundary

$$\hat{q} = q(\{\beta\},B), \quad (7.4)$$

let alone the implicit dependence on the models of the EAS absorption, atmosphere mass thickness and resolution function, including discrepancy between the true EAS arrival direction and direction of the front's local tangent plane.

The estimation of the dispersion value of the solution (7.4) of the maximal likelihood equation is defined by the well known [17] relation:

$$\sigma_q^2(\{\beta\},B) = -\left[\frac{d^2}{dq^2}\mathfrak{L}(q|\{\beta\},B)\right]_{q=\hat{q}}^{-1}.$$

The second-order derivative is explicitly available due to the known form (6.2) of the bounded distorted distribution:

$$\frac{d^2}{dq^2}\mathfrak{L}(q|\{\beta\},B) =$$
$$= -N_B\left\{\begin{matrix}\left[\mathrm{Vi}_2(q,B) - \mathrm{Vi}_1(q,B)^2\right] - \\ -\frac{1}{N_B}\sum_{j=1}^{N_B}\left[\mathrm{Vr}_2(q,\beta_j) - \mathrm{Vr}_1(q,\beta_j)^2\right]\end{matrix}\right\} \quad (7.5)$$

Here the average functions of the squares of the normalized air thickness function $V(\alpha)$ (4.3) are used:

$$\mathrm{Vr}_2(q|\beta) = \frac{\int_0^1 \mathrm{Rr}(\beta|\xi)\cdot f_\alpha(\xi|q)\cdot V(\xi)^2\,d\xi}{\int_0^1 \mathrm{Rr}(\beta|\xi)\cdot f_\alpha(\xi|q)\,d\xi}; \quad (7.6)$$

$$\mathrm{Vi}_2(q|B) = \frac{\int_0^1 \mathrm{Ri}(B|\xi)\cdot f_\alpha(\xi|q)\cdot V(\xi)^2\,d\xi}{\int_0^1 \mathrm{Ri}(B|\xi)\cdot f_\alpha(\xi|q)\,d\xi}. \quad (7.7)$$

The expression for the second-order derivative of the logarithm of the likelihood function consists of two types of functions: the sample-mean part of the averages (7.6) and (7.2) in the second line of the expression (7.5) and the sample-independent part (7.7) and (7.3) in the first line. So the required dispersion of the $q$ value estimation can be calculated immediately by summation of quadratures, without unstable numerical differentiation.

## 8. EAS absorption path estimations for several trimming boundaries of the TSU data sample and final estimation of the absorption path required

Here we at last begin the estimation of the EAS absorption path $\Lambda$. At the first stage we estimate the values of the parameter $q$ for the used set of arbitrary trimming boundaries $B_k$:

$$B_k = 0.300 + 0.025\cdot k; \quad k = 0,1,\ldots 28; \quad B_k \in [0.3, 1.0].$$

The estimations are obtained by means of the numerical solutions of the corresponding maximal likelihood equation (7.1). Hereinafter all numerical calculations are applied to the only TSU data sample $\{\beta\}_{\mathrm{TSU}}$ (order statistics); that is why the sample references are omitted. The spheric atmosphere model ((3.3),(3.4)) is used.

The set of the estimated vertical absorption range number parameters $\hat{q}_k = q(B_k)$ (with the appropriate standard deviations) are shown in Figure 6.

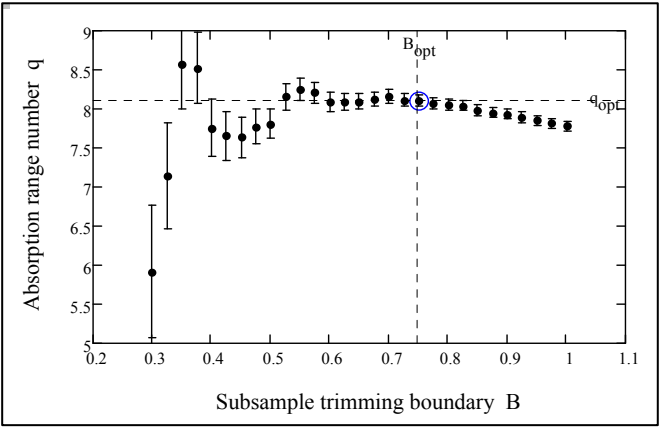

**Figure 6** The vertical absorption range number parameter q dependence on the trimming boundary B

Note the unstable behavior of the $\hat{q}_k$ estimations dependent on the trimming boundary $B$ at the left part of the graph. Obviously the estimations upon the low-boundary subsamples are impressively unsafe.

The appropriate estimations of the EAS absorption path $\hat{\Lambda}_k$ are calculated under the definition (4.3):

$$\hat{\Lambda}_k = X^\uparrow_{\text{TSU}} / \hat{q}_k .$$

The set of the estimated EAS absorption paths are shown in Figure 7. It should be noted that every estimation point in this figure is statistically dependent on the previous one (at a lesser trimming boundary), as each completely uses the previous subsample for the current estimation.

In Figure 7 the segment of approximate 1σ-stability of the $\Lambda_k$ estimations is marked out. At the final stage the optimal resulting values of the absorption range number and absorption path are accepted to be

$$\hat{q}_{\text{opt}} = 8.11; \qquad \hat{\sigma}_q = 0.08;$$
$$\hat{\Lambda}_{\text{opt}} = 120.7\, g/cm^2; \qquad \hat{\sigma}_\Lambda = 1.2\, g/cm^2 . \qquad (8.1)$$

by use of the stability condition in the *B*-segment defined by the requirement

$$\forall \{\hat{\Lambda}_k\} \in \left[ \hat{\Lambda}_{\text{opt}} - \hat{\sigma}_\Lambda,\ \hat{\Lambda}_{\text{opt}} + \hat{\sigma}_\Lambda \right].$$

This segment is quite wide: $0.585 \leq B \leq 0.825$ and approximately corresponds (under the assumption $\alpha = \beta$) to the allowed angular boundaries in the segment $36° \leq \theta \leq 56°$.

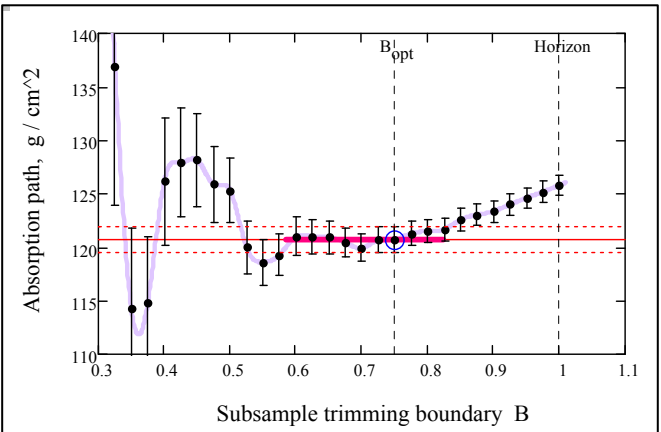

**Figure 7** The absorption path Λ estimation dependence on the subsample trimming boundary *B*.
The point-connecting curve is shown for a vision convenience.
The segment of 1σ-stability is marked out by the bold line.

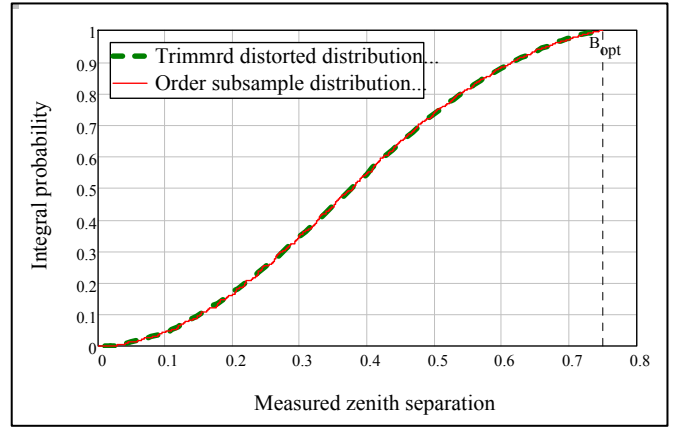

**Figure 8** The optimal order subsample distribution comparison with the respective distorted integral distribution.

The maximal absolute difference $D^\pm_{N_{0.75}} = 9.53 \cdot 10^{-3}$ between the order statistics distribution of the observed separations and the respective trimmed distorted distribution (Figure 8) upon the optimal subset $0 \leq \beta < B_{\text{opt}} = 0.75$ containing $N_{0.75} = 20150$ events indicates that the observation probability of the lesser difference is only 2.6% according to the Kolmogorov criterion [18].

## Discussion

The measurement errors broaden the distribution of the existing data compared to the corresponding fundamental physical distribution.

The main influence of this distorting feature is explicitly expressed in the difference of the fundamental distribution (4.4) and the fitted distorted one (6.2), shown in Figure 9. Any precarious attempt to fit the fundamental distribution to the existing data results in an unstable estimation of the EAS absorption path, not in agreement with existing world data.

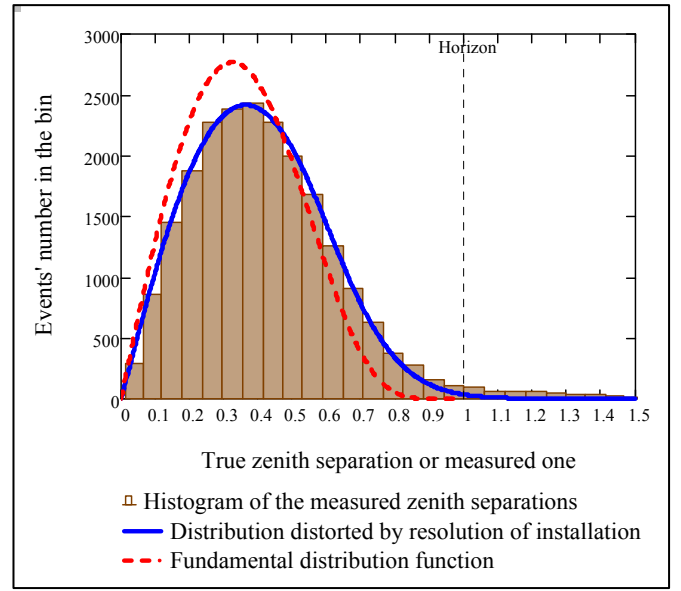

**Figure 9** Two distributions of zenith separations by the TSU data
The optimized distorted distribution (6.2) of the measured separations is compared with the events' number histogram by the TSU data. The correspondent distribution of true zenith separations (4.4) is shown too.
The distributions are normalized to the histogram containing 21684 EAS events' data.

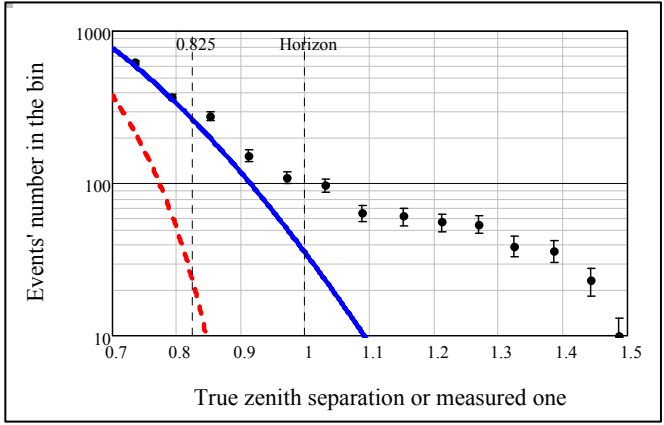

**Figure 10** The same comparison as in **Figure 9** shown in the horizon vicinity α = 1. The key symbols are the same too, but the histogram is shown by the points with errors. The growing relative difference between data and optimal distorted distribution at the values of measured zenith separations exceeding the upper boundary B = 0.825 of the Λ-estimation stability is clearly visible.

The TSU data in the horizon vicinity, shown in Figure 10, display the steady excess over the curve predicted by the distorted distribution (6.2) $f_\beta(\beta|\hat{q}_{opt}, 1.5)$ with the optimal (8.1) EAS absorption path. This behavior is unexpected and unexplained. It looks like a broad band of substance deficiency near the horizon in comparison with the spheric atmosphere model, which itself underestimates (Figure 3) the air mass thickness in the horizon limit. Normally the matter excess is expected here. The known behavior of the equivalent dispersion (Figure 4) does not allow a sufficient broadening of the resolution function.

## Conclusions

It has been established by investigation of the TSU data that accounting for the resolution function specific for the TSU installation makes it possible to validate the fundamental distribution (4.4) of the true zenith separations $\alpha = \sin(\theta)$. This conventional model of EAS absorption, in accordance with the spheric layer atmosphere model approximation (3.3), has proved to be valid for the description of the EAS absorption process within the interval $0 \leq \beta \leq 0.825$ of the measured zenith separation $\beta$, i.e. in the interval $0 \leq \theta \leq 56°$ of the zenith angle (under the assumption $\alpha = \beta$).

The estimated value of the EAS absorption path is actually stable under variation of upper limits of the $\beta$-value data trimming within the $0.525 \leq B \leq 0.825$ segment. Any estimation of this parameter upon the more restricted sequence of intervals of $\beta$ variable is unstable.

It is the immediate consequence of this study that any attempt to estimate the EAS absorption path with use of some data trimming, not proved to be consistent with stability under variation of this trimming limit, is unreliable. Our $\Lambda_{TSU} = (120.7 \pm 1.2)\,g/cm^2$ estimation is in approximate agreement with the previous estimations by installations located at various altitudes:

$\Lambda_{Gr} = (135 \pm 10)\,g/cm^2$; [1]

$\Lambda_{T\text{-}S} = (130 \pm 7)\,g/cm^2$; [3]

$\Lambda_{LAAS} = (106 \pm 6)\,g/cm^2$; [5]

$\Lambda_{MSU} = (115 \pm 4)\,g/cm^2$; [6]

$\Lambda_{TBS} = (115.4 \pm 2.6)\,g/cm^2$; [7]

$\Lambda_{TEL} = (131.1 \pm 1.4)\,g/cm^2$. [8]

## *Acknowledgements*

The authors are grateful to other current and former members of our group for their technical support. We are especially thankful to our colleagues working now in foreign. This work was supported by the Georgian National Science Foundation subsidy for a grant of scientific researches #GNSF/ST06/4-075 (№ 356/07) and by Shota Rustaveli National Science Foundation, Project #DI/6/6-300/12.